\documentclass[
 reprint,
 amsmath,amssymb,
 aps,
 prd,
 preprintnumbers
]{revtex4-1}

\usepackage[utf8x]{inputenc}
\SetUnicodeOption{mathletters}
\SetUnicodeOption{autogenerated}

\usepackage{xcolor}
\usepackage[colorlinks=true,
            linkcolor={red!50!black},
            urlcolor={green!50!black},
            citecolor={blue!80!black}]{hyperref}

\usepackage{amsthm}
\usepackage{amsfonts}
\usepackage{mathrsfs}
\usepackage{bm}
\usepackage{tensor}

\newcommand{\mr}{\ensuremath{\mathrm}}
\newcommand{\mc}{\ensuremath{\mathcal}}
\newcommand{\ms}{\ensuremath{\mathsf}}
\newcommand{\mb}{\ensuremath{\mathbb}}
\newcommand{\mh}{\ensuremath{\mathscr}}
\renewcommand{\v}[1]{\ensuremath{\bm{\mathbf{#1}}}}
\renewcommand{\d}{\mr{d}}
\newcommand{\R}{\mb{R}}	

\renewcommand{\l}{\left}
\renewcommand{\r}{\right}
\newcommand{\f}{\frac}
\newcommand{\tf}{\tfrac}
\newcommand{\s}{\sqrt}
\newcommand{\h}{\hat}
\newcommand{\I}{\indices}

\renewcommand{\Re}{\operatorname{Re}}
\DeclareMathOperator{\tr}{tr}

\begin{document}

\preprint{QMUL-PH-15-20}

\title{Global equilibrium and local thermodynamics\\ in stationary spacetimes}

\author{Rodolfo Panerai}
\email{r.panerai@qmul.ac.uk}
\affiliation{
 Centre for Research in String Theory \\
 School of Physics and Astronomy \\
 Queen Mary University of London \\
 Mile End Road, London E1 4NS, United Kingdom
}

\begin{abstract}
In stationary spacetimes global equilibrium states can be defined, applying the maximum entropy principle, by the introduction of local thermodynamic fields determined solely by geometry. As an example, we study a class of equilibrium states for a scalar field in the Einstein's static universe, characterized by inhomogeneous thermodynamic properties and non-vanishing angular momentum.
\end{abstract}


\maketitle

\section{Introduction}
The concept of local temperature in quantum field theory has been the subject of various studies in recent years \cite{Buchholz2002, Buchholz2006, Solveen2012, Becattini2014, Buchholz2015}. This has its relevance, not only in characterizing systems out of equilibrium, but also when dealing with the case of curved spacetime, where even a condition of global equilibrium is associated to inhomogeneous thermodynamic properties \cite{Tolman1930}.

It is well known that stationary spacetimes admit states of global equilibrium. This can be obvious from physical intuition, as one expects, for a thermal system to maintain its condition of equilibrium, the existence of a frame in which geometry does not depend on the time coordinate.
Yet, from the point of view of statistical mechanics, the presence of a globally timelike Killing vector field implies the notion of a conserved energy, which is usually a starting point in the construction of statistical ensembles of global equilibrium.

This point of view, however, hinders an interpretation of equilibrium based on \emph{local} quantities. In fact, while global aspects are crucial in ensuring for the thermal description to be compatible with the unitary evolution of the quantum theory, one may want to characterize a thermal system in terms of local thermodynamic properties an observer can measure with local experiments.

Our approach will be exactly this: starting from local aspects and moving up to global ones. Through a natural generalization of the \emph{maximum entropy principle}---a principle which lies at the heart of equilibrium statistical mechanics---we will enforce a condition of equilibrium based on local constraints; then, maintaining an interpretation based on this description, we will recover global equilibrium by imposing unitarity.

This picture has a close relation to the ones proposed in \cite{Zubarev1996, vanWeert1982, Becattini2014} although our aim is not to perform a hydrodynamic expansion of the quantum theory but rather to fully characterize global equilibrium states in terms of their local thermodynamics. Each state is in 1-1 correspondence with a timelike Killing vector field, as this contains the complete information on thermodynamic variables.

As an example of this, we will study inhomogeneous thermal states of a scalar field in the \emph{Einstein's static universe}. This spacetime admits a continuous class of equilibrium states describing a rigidly rotating thermal system with different values of angular momentum. This is not to be intended as a cosmological model, as in our example the spacetime will be regarded as a fixed background.

\subsection*{Notation and conventions}
We adopt natural units, i.e.\ we set $c=\hslash=k_{\mr{B}}=1$. The metric tensor has the ``mostly minus'' signature; the Riemann and the Ricci tensor are defined, respectively, as $R\I{^{\alpha}_{\beta\gamma\delta}} = \partial \I{_{\delta}} \Gamma \I{^{\alpha}_{\beta\gamma}} - \ldots$ and $R\I{_{\alpha\beta}} = R\I{^{\gamma}_{\alpha\gamma\beta}}$. The Levi-Civita tensor is defined in terms of the the Levi-Civita symbol $\tilde{\epsilon}^{\mu\nu\rho\sigma}=[\mu\nu\rho\sigma]$ as $\epsilon^{\mu\nu\rho\sigma} = [-g]^{-\f{1}{2}}\tilde{\epsilon}^{\mu\nu\rho\sigma}$.

\section{Generalizing equilibrium}\label{SEC:generalizing_equilibrium}
In equilibrium quantum statistical mechanics a physical state is described in terms of few macroscopic parameters. In particular one can choose a set of \emph{relevant observables} $\{A_i\}$ and define the equilibrium state by fixing the corresponding expectation values to a certain set $\{\overline{A_i}\}$. Among all the different density operators that fill this requirement the equilibrium operator is selected, according to the maximum entropy principle \cite{Jaynes1957-1, *Jaynes1957-2}, as the one that maximizes the \emph{von Neumann entropy} functional $S_{\mr{VN}}[\varrho] = -\tr(\varrho \log \varrho)$. This can be achieved by introducing a set of Lagrange multipliers $\{\lambda_i\}$ and the result is the well-known Gibbs-like density operator
\begin{equation}
\varrho = Z^{-1}(\lambda_i) \, \exp\!\Big(-\sum_i \lambda_i A_i\Big) \, ,
\end{equation}
where the partition function $Z(\lambda_i)$ is defined so that $\varrho$ is normalized. The von Neumann entropy then reads
\begin{equation}
S_{\mr{VN}} = \log Z + \sum_i \lambda_i \overline{A_i} \, .
\end{equation}

At this point one could find the relation that holds between the Lagrange multipliers $\lambda_i$ and the expectation values $\overline{A_i}$ and use it to express $\varrho$ as a function of the latter, however it is customary to maintain the dependency of the statistical operator from the multipliers $\lambda_i$ which are then interpreted as statistical parameters.

In the simple case where the $A_i$ are conserved quantities the state is actually static, i.e.\ in Schr\"odinger representation $\dot{\varrho}=0$. A well-known example of this is the canonical ensemble defined by
\begin{equation}\label{rho_canonical}
\varrho_{\mr{can.}}^{\vphantom{-1}} = Z_{\mr{can.}}^{-1}(\beta) \, \exp(-\beta H) \, .
\end{equation}

In special-relativistic theories where a unitary representation $U(\Lambda,a)$ of the Poincar\'e group is defined, one looks for forms of the density operator which are covariant, 
meaning that every inertial observer can describe the same system using the same density operator $\varrho$, simply by changing the values $\lambda_i$.

This is not always the case, as can be seen for the canonical ensemble (\ref{rho_canonical}). The problem is that under a boost $\varrho_{\mr{can.}}^{\vphantom{-1}}$ acquires a dependence on the spatial components of the four-momentum $P$, but in (\ref{rho_canonical}) only $P^0$ appears. We need then to include all the components of the irreducible representation of the Lorentz group to which $P^0$ belongs; this leads to the \emph{covariant canonical ensemble}
\begin{equation}\label{rho_canonical_covariant}
\varrho_\beta = Z^{-1}(\beta)\,\exp(-\beta_\mu P^\mu) \, .
\end{equation}
Now, because
\begin{equation}
U(\Lambda,a) \, \varrho_\beta \, U^{\dagger}(\Lambda,a) = \varrho_{\Lambda\beta} \,,
\end{equation}
$\beta$ is actually a four-vector, and since $\langle P\rangle_\beta \parallel \beta$ we know that it has to be time-like. In this way $\beta$ defines a four-velocity unit vector $u = [\beta^2]^{-\f{1}{2}} \beta$ and identifies a rest frame in which $u^i = 0$. Proper temperature is defined as the Lorentz scalar $T = [\beta^2]^{-\f{1}{2}}$.

Notice, however, how in Minkowski spacetime $\varrho_\beta$ is technically not well defined as $e^{-\beta\cdot P}$ is not a trace-class operator since $H$ has a continuous spectrum \cite{Haag1992}. A condition of equilibrium can still be defined through the thermodynamic limit of the system quantized in a finite volume, i.e.\ on the flat spacetime $(\R\times \mb{T}^3, \eta)$ where a conserved four-vector $P$ is defined.

In the following discussion we will restrict ourself to those cases in which the spectrum of the theory allows for a well-defined density operator.

\subsection{Inhomogeneous equilibrium}
Since (\ref{rho_canonical_covariant}) is invariant under spacetime translations, equilibrium states described by the covariant canonical ensemble are homogeneous: the expectation value $\langle O(x)\rangle_\beta$ of any local observable is independent of the spacetime point $x$ in which it is evaluated. To describe a generic physical state whose statistical properties varies in space and time one has to choose the relevant observables among local ones \citep{Zubarev1996}.

In the context of a relativistic quantum theory, a set $\{A_i\}$ of relevant conserved charges defines a state of global equilibrium. We know that for each $A_i$, a conserved current $j_i$ exists, such that
\begin{equation}
A_i = \int_\sigma \d \sigma \, n_\mu j_i^\mu \,,
\end{equation}
where $\sigma$ is an arbitrary Cauchy surface, and $n$ its normal unit four-vector. It is natural then to define a local equilibrium density operator \cite{vanWeert1982} with
\begin{equation}\label{rho_LE_generic}
\varrho[\lambda_i] = Z^{-1}[\lambda_i] \, \exp\!\Big(-\int_\sigma \d \sigma \, n_\mu \sum_i \lambda_i^{\vphantom{\mu}} j_i^\mu \Big) \, ,
\end{equation}
where the choices of $\sigma$ and of the smooth function $\lambda_i|_\sigma$ define the initial conditions.

We want to stress that, although this is not the only way to define local equilibrium, (\ref{rho_LE_generic}) is the natural choice if one wants a local equilibrium density operator that obeys the maximum entropy principle: $\varrho[\lambda_i]$, in fact, satisfies a continuum of constraints associated to the expectation values of a continuous set of local observables $j_i(x)$.

We will now briefly turn our attention to a key aspect of the expression (\ref{rho_LE_generic}) concerning time evolution. This can be more easily discussed by restricting our analysis to constant-time Cauchy surfaces. While it is true that in Heisenberg representation density operators do not evolve, on the other hand one may want to apply time evolution to $\varrho$ to describe the local equilibrium state in terms of the evolution of the fields $\lambda_i$. Again, this cannot be done in general, because the unitary time translation would not preserve the form of the density operator. In other words, if we call $T(t_0,t)$ the unitary time-evolution operator, there is no $\lambda_i(t,\v{x})$ such that
\begin{equation}\label{rho_LE_generic_evolution}
T(t_0,t) \, \varrho[\lambda_i(t_0,\v{x})] \, T^{\dagger}(t_0,t) = \varrho[\lambda_i(t,\v{x})] \, .
\end{equation}

Moreover, it is easy to see that a unitary evolution implies that the von Neumann entropy remains constant in time while one, in general, would expect it to increase. This means that with unitary evolution information is conserved, and in the local equilibrium case there is a flow of information from relevant observables to irrelevant ones. One way to deal with this is to modify the time evolution by the introduction of a projection operator that would destroy irrelevant information keeping the density operator of the form (\ref{rho_LE_generic}) \cite{Balian1999}.

However, one may ask if, apart from the constant ones, particular choices of $\lambda_i$ fields exist such that the thermal evolution of the corresponding states coincides with the unitary evolution prescribed by the quantum theory. Anti-evolving both sides of (\ref{rho_LE_generic_evolution}), we see that this is the case if (\ref{rho_LE_generic}) is independent of the choice of the Cauchy surface $\sigma$.

We want to focus our attention in particular on the generalization to local equilibrium of the covariant canonical statistical operator (\ref{rho_canonical_covariant}), that is
\begin{equation}\label{rho_LT}
\varrho_{\mr{LT}}^{\vphantom{-1}}[\beta] = Z_{\mr{LT}}^{-1}[\beta] \, \exp\!\Big( -\int_\sigma \d \sigma \, n_\mu T^{\mu\nu} \beta_\nu \Big) \,,
\end{equation}
and look for a time-like $\beta$ field such that $\varrho_{\mr{LT}}$ is independent of the choice of $\sigma$. This condition is satisfied if $\beta$ is a Killing vector field. If (\ref{rho_LE_generic_evolution}) holds, then the von Neumann entropy coincides with the thermodynamic entropy which remains constant, so the state is actually a state of global equilibrium.

Notice that (\ref{rho_LT}) is written in a general covariant fashion and can be applied also for curved spacetimes where a conserved four-momentum is not defined. A generic spacetime that admits a global time-like Killing vector field $\xi$ is called \emph{stationary}. In such a spacetime one can define a conserved quantity
\begin{equation}
Q = \int_\sigma \d \sigma \, n_\mu T^{\mu\nu} \xi_\nu
\end{equation}
and a state of global equilibrium with $\beta \propto \xi$ which leads to a density operator of type (\ref{rho_canonical}): it is a general feature, even for curved-space, that thermodynamic equilibrium needs a notion of \emph{time}. The difference with the flat-space homogeneous case is that now the proper temperature $T$ is not uniform while the expression $T[\xi^2]^{\f{1}{2}}$ is: this is the \emph{Tolman-Ehrenfest effect} \cite{Tolman1930}.

Notice how one could, in principle, introduce a different definition of temperature, where this plays the role of an overall constant multiplying $Q$. That definition is of little value in our approach.

\subsection{Stress-energy tensor}
In the previous paragraph we noticed how a globally time-like Killing vector field $\beta$ defines, through (\ref{rho_LT}), a state of global equilibrium with inhomogeneous physical properties. We pointed out that $\beta$ canonically identifies two local quantities, namely $T$ and $u$, that play the role of thermodynamic parameters. However,  while we know that $T$ and $u$ are in some way related to the description of local thermal properties, they are, as for now, associated mainly with a statistical interpretation and we have yet to derive their relation with any physical observable.

A fundamental observable in the description of thermal systems is the stress-energy tensor, indeed fixing its expectation value was the starting point of our analysis. In the case of homogeneous equilibrium in flat space its thermal expectation value takes the well-known \emph{ideal form}
\begin{equation}\label{ideal_T}
T^{\mu\nu} = (\epsilon+p) u^\mu u^\nu - p g^{\mu\nu} \,, \quad \epsilon=\epsilon(T), \quad p=p(T) \,,
\end{equation}
whose simple tensor structure is strictly determined by the symmetries of (\ref{rho_canonical_covariant}). On the other hand, in the more general case we are dealing with, the equilibrium state may not have the same symmetries and $\langle T^{\mu\nu}\rangle_{\mr{LT}}$ may not be an ideal stress-energy tensor. This is in sharp contrast to the general belief according to which, in absence of dissipative phenomena, (\ref{ideal_T}) does hold.

It is true that the concept of local thermalization itself is usually associated with the existence of a microscopic scale at which individual regions of the system appear at homogeneous equilibrium, meaning that the typical thermal correlation lengths, which determine the distance beyond which a point cannot ``see'' remote thermal properties, are much smaller than the distance over which $\beta$ varies. In this regime, corrections to the ideal form can be smaller than thermal fluctuations. The density operator (\ref{rho_LT}), however, is more general and allows for equilibrium states characterized by strong $\beta$ gradients where the scale separation described above is no longer valid.

The von Neumann entropy for a state of equilibrium characterized by a Killing field $\beta$ reads
\begin{equation}
S_{\mr{VN}} = \log Z_{\mr{LT}} - \int_\sigma \d \sigma \, n_\mu \langle T^{\mu\nu}\rangle_{\mr{LT}} \beta_\nu \,,
\end{equation}
which, being extensive, diverges in the thermodynamic limit. Now, since $\log Z_{\mr{LT}}$ is a conserved quantity, for a certain $\zeta>0$ holds
\begin{equation}\label{zeta_def}
\log Z_{\mr{LT}} = \zeta \int_\sigma \d \sigma \, n_\mu \beta^\mu \,,
\end{equation}
and one can define a conserved entropy density four-vector
\begin{equation}\label{s_def}
s^\mu = (\zeta g^{\mu\nu} - \langle T^{\mu\nu}\rangle_{\mr{LT}}) \beta_\nu 
\end{equation}
such that
\begin{equation}
S_{\mr{VN}} = \int_\sigma \d \sigma \, n_\mu s^\mu \,.
\end{equation}

\section{Einstein's static universe}\label{SEC:ESU}
In Minkowski spacetime, globally time-like Killing vector fields define only states of homogeneous equilibrium. The Einstein's static universe (ESU), on the contrary, admits a class of inhomogeneous states, describing a spinning thermal system, which can be obtained from a smooth deformation of the canonical homogeneous equilibrium state.

We want to stress that, as mentioned in the Introduction, here the geometry of the spacetime plays the role of a fixed background, which means that we do not impose for the thermal average of the stress energy tensor to satisfy the Einstein's equations. After all, as we will see, symmetries associated with the inhomogeneous states constitute in general only a subgroup of the isometries of the spacetime.

\subsection{Geometry}
For the ESU $(\R\times \mb{S}^3,g)$ we adopt ultra-static coordinates $\{t,\chi,\theta,\varphi \}$, where $t\in \R$ is time and $\chi,\theta\in[0,\pi]$, $\varphi\in[0,2\pi)$ are hyperspherical coordinates on the 3-sphere. In terms of these coordinates the metric tensor reads
\begin{equation}
g = \d t^2 - a^2 \l[ \d \chi^2 + \sin^2\!\chi \l( \d \theta^2 + \sin^2\!\theta \, \d \varphi^2 \r) \r] \,,
\end{equation}
where $a$ is the radius of the universe. The non-vanishing components of the Ricci tensor are
\begin{align}
R_{\chi\chi} &= -2 \,, \nonumber \\
R_{\theta\theta} &= -2\sin^2\!\chi \,, \\
R_{\varphi\varphi} &= -2\sin^2\!\chi \sin^2\!\theta \,, \nonumber 
\end{align}
and the scalar curvature is $R=6a^{-2}$.

Isometries correspond to time translations and rotations on the 3-sphere, which together form the group $\ms{T}^1\times \ms{SO}(4)$. While time translations are obviously generated by the Killing field $\v{e}_t$, generators of rotations can exhibit a less simple expression in our choice of coordinates, depending on the axis about which the rotation is performed. Without loss of generality we consider the rotations generated by $\v{e}_\varphi$, which correspond to the transformations $\varphi\mapsto\varphi+\alpha$. We then define our inverse temperature four-vector as a linear combination of $\v{e}_t$ and $\v{e}_\varphi$, and write it in terms of two global parameters, $\tau>0$ and $|\omega |<1$, as
\begin{equation}\label{beta_ESU}
\beta^\mu=\tau^{-1}(1,0,0,a^{-1}\omega) \,.
\end{equation}
This, in turn, defines a proper temperature field
\begin{equation}
T(\chi,\theta) = \f{\tau}{\s{1 - \omega^2 \sin^2\!\chi \sin^2\!\theta}}
\end{equation}
and a four-velocity field
\begin{equation}\label{u_field}
u^\mu = \f{1}{\s{1 - \omega^2 \sin^2\!\chi \sin^2\!\theta}} \, (1, 0, 0, a^{-1}\omega) \,.
\end{equation}

When $\omega\neq 0$ the integral curves of $u$ are, in general, not geodesic. Indeed, the four-acceleration field reads
\begin{equation}
A^\mu = -\f{\omega^2 (0, \sin\chi \cos\chi \sin^2\!\theta, \sin\theta \cos\theta, 0)}{a^2 (1 - \omega^2 \sin^2\!\chi \sin^2\!\theta)} \,.
\end{equation}

The picture that one gets from the $u$ and $T$ fields is that of a thermal system that rotates rigidly on the 3-sphere with a local proper temperature that varies in the range
\begin{equation}
\tau \leq T \leq \f{\tau}{\s{1-\omega^2}} \,,
\end{equation}
reaching the minimum at ``poles'', and the maximum at the ``equator''.

Our choice of $\beta$ breaks the symmetry group given by isometries leaving a residual $\ms{T}^1 \times \ms{SO}(2) \times \ms{SO}(2)$ symmetry generated by $\v{e}_t$, $\v{e}_\varphi$, and $r^\mu = (0, -\cos\theta, \cot\chi \sin\theta, 0)$. In particular, knowing that the properties of the system do not vary along the orbits generated by $r$ will allow us to simplify some of our results performing the rotation $(t,\chi,\theta,\varphi)\mapsto(t,\alpha,\f{\pi}{2},\varphi)$, for $\alpha(\chi,\theta)$ such that
\begin{equation}\label{alpha_eq}
\sin\alpha = \sin\chi \sin\theta 
\end{equation}
(see Appendix \ref{APP:residual_symmetries}).

\subsection{Scalar field quantization}
Let us consider a scalar field described by the lagrangian density
\begin{equation}\label{KG_Lagrangian}
\mh{L}[\Phi] = \tf{1}{2}\s{-g} \big[g^{\mu\nu} \nabla_{\mu}\Phi \nabla_{\nu}\Phi - (M^2 + \xi R) \Phi^2\big] \,,
\end{equation}
which determines the Klein-Gordon equation
\begin{equation}\label{KG_eq}
\big( \square_g + M^2 + \xi R \big) \Phi = 0 \,,
\end{equation}
where $\square_g = g^{\mu\nu}\nabla_{\mu}\nabla_{\nu}$ is the D'Alambert operator in the ESU. A complete set of orthonormal solutions of \eqref{KG_eq} is given by \cite{Ford1976}
\begin{equation}
u_{n,\ell,m}(x) = \f{1}{\s{2a^3\omega_n}} \, e^{-i \omega_n t} \, \Pi^\ell_n(\chi) \, Y^m_\ell(\theta,\varphi)
\end{equation}
where $n\in \mb{N}$, $\ell\in \{0, \ldots, n\}$, $m\in \{-\ell, \ldots, \ell \}$, and
\begin{equation}
\omega_n = \s{a^{-2}n(n+2)+M^2+\xi R} \,.
\end{equation}
The $Y^m_\ell(\theta,\varphi)$ are the usual spherical harmonics, while $\Pi^\ell_n(\chi)$ are given by
\begin{equation}
\Pi^\ell_n(\chi) = \f{ \ell! \, 2^{\ell+\f{1}{2}} \s{(n-l)!(n+1)} }{ \s{\pi(n+l+1)!} } \, \sin^\ell \!\chi \, C^{\ell+1}_{n-\ell}(\cos \chi) \,,
\end{equation}
where $C^\lambda_n(x)$ are Gegenbauer polynomials. More details can be found in Appendix \ref{APP:orthonormality}.

The special choice of parameters $M=0$, $\xi=\f{1}{6}$ constitutes the so-called \emph{conformal coupling}. In that case the spectrum of the theory becomes particularly simple, as $\omega_n=a^{-1}(n+1)$. Although our approach is not inherently restricted to the conformal case, we will derive our end results only for this particular choice of parameters as this greatly simplifies calculations.

The field operator can be written in terms of the solutions of the equation of motion as
\begin{equation}
\Phi(x) = \sum_{n=0}^\infty \sum_{\ell=0}^n \sum_{m=-\ell}^{+\ell} \l(u^{\vphantom{\dagger}}_{n,\ell,m}(x)\,a^{\vphantom{\dagger}}_{n,\ell,m} + \mr{h.c.} \r) \,,
\end{equation}
where creation and destruction operators obey the usual commutation relations
\begin{equation}
[a^{\vphantom{\dagger}}_{n,\ell,m}, a^{\dagger}_{n',\ell',m'}] = \delta^{\vphantom{\dagger}}_{n,n'} \delta^{\vphantom{\dagger}}_{\ell,\ell'} \delta^{\vphantom{\dagger}}_{m,m'} \,.
\end{equation}

Varying \eqref{KG_Lagrangian} with respect to the metric one obtains the stress energy tensor
\begin{multline}\label{stress_energy_tensor}
T_{\mu\nu} = \nabla_\mu\Phi \, \nabla_\nu\Phi - \tf{1}{2} g_{\mu\nu} (g^{\rho\sigma} \nabla_\rho\Phi \, \nabla_\sigma\Phi - M^2\Phi^2) \\ + \xi(g_{\mu\nu}\square_g - \nabla_\mu\nabla_\nu - G_{\mu\nu}) \Phi^2 \,.
\end{multline}
We are now ready to define a density operator of the type \eqref{rho_LT} with inverse temperature four-vector field \eqref{beta_ESU}. Taking a constant-time Cauchy surface $\sigma_t$ with associated volume form $\d \sigma_t =  a^3 \sin^2\!\chi \sin\theta \, \d \chi \wedge \d \theta \wedge \d \varphi$ we can write the density operator as
\begin{equation}
\varrho_{\mr{LT}}^{\vphantom{-1}} =  Z_{\mr{LT}}^{-1} \exp\!\l(-\tau^{-1} \l[H - a^{-1} \omega J\r]\r) \,,
\end{equation}
in terms of energy and angular momentum operators
\begin{equation}
H = \int_{\sigma_t} \d \sigma_t \; T_{tt} \,, \qquad J = -\int_{\sigma_t} \d \sigma_t \; T_{t\varphi} \,,
\end{equation}
where both quantities, being conserved, do not depend on $t$. It is straightforward to show that
\begin{equation}
\begin{aligned}
H &= \sum_{n,\ell,m} \omega_n \big(a^{\dagger}_{n,\ell,m}a^{\vphantom{\dagger}}_{n,\ell,m} + \tf{1}{2}\big) \,, \\
J &= \sum_{n,\ell,m} m \, a^{\dagger}_{n,\ell,m}a^{\vphantom{\dagger}}_{n,\ell,m} \,.
\end{aligned}
\end{equation}

\subsection{Partition function}
Since $H$ and $J$ are both diagonal in the occupation number basis, the partition function can be easily computed as
\begin{align}
Z_{\mr{LT}} &= \sum_{\{N_{n,\ell,m}\}} \langle \{N_{n,\ell,m}\}| \exp(-\kappa[aH-\omega J]) |\{N_{n,\ell,m}\}\rangle \nonumber \\
  &= \prod_{n,\ell,m} \sum_{N=0}^{\infty} \exp(-\kappa[a\omega_n-\omega m])^N \,,
\end{align}
where for convenience we introduced the adimensional parameter $\kappa=(a\tau)^{-1}$ and neglected any contribution from the vacuum energy as it amounts to a rescaling of the partition function which does not affect the physics of the system.

Taking the logarithm we find
\begin{equation}
\begin{aligned}
\log Z_{\mr{LT}} &= - \sum_{n,\ell,m} \log(1-\exp(-\kappa[a\omega_n-\omega m])) \\
  &= \sum_{n,\ell,m} \sum_{j=1}^{\infty} j^{-1} \exp(-j\kappa[a\omega_n-\omega m]) \,.
\end{aligned}
\end{equation}
Now, since
\begin{equation}
\sum_{\ell=0}^n \sum_{m=-\ell}^\ell e^{mz} = \f{\sinh^2(\tf{1}{2}[n+1]z)}{\sinh^2(\tf{1}{2}z)} \,,
\end{equation}
then
\begin{equation}
\log Z_{\mr{LT}} = \sum_{j=1}^{\infty} j^{-1} \sum_{n=0}^{\infty} e^{-j\kappa a\omega_n} \, \f{\sinh^2(\tf{1}{2}[n+1]j\kappa\omega)}{\sinh^2(\tf{1}{2}j\kappa\omega)} \,.
\end{equation}

For the conformally coupled theory the previous expression reduces to
\begin{equation}
\begin{aligned}
\log Z_{\mr{LT}} &= \sum_{j=1}^{\infty} \f{1}{j \sinh^{2}(\tf{1}{2}j\kappa\omega)} \sum_{n=1}^{\infty} e^{-nj\kappa} \sinh^2(\tf{1}{2}nj\kappa\omega) \\
  &= \sum_{j=1}^{\infty} \f{\coth(\tf{1}{2}j\kappa)}{4j \sinh(\tf{1}{2}j\kappa[1+\omega]) \sinh(\tf{1}{2}j\kappa[1-\omega])} \,.
\end{aligned}
\end{equation}

By taking derivatives of $\log Z_{\mr{LT}}$ one can extract the expectation value of $H$ and $J$. In particular the latter results in the compact expression
\begin{equation}
\langle J\rangle_{\mr{LT}} = \sum_{j=1}^{\infty} \f{\coth(\tf{1}{2}j\kappa) \sinh(j\kappa\omega)}{2[\cosh(j\kappa) - \cosh(j\kappa\omega)]^2} \,.
\end{equation}

Knowing $\log Z_{\mr{LT}}$, it is straightforward to compute the associated \emph{Renyi entropy}
\begin{equation}
\begin{aligned}
S_r &= \f{1}{1-r} \log \tr \varrho_{\mr{LT}}^r \\
  &= \f{\log Z_{\mr{LT}}[r\beta] - r \log Z_{\mr{LT}}[\beta]}{1-r} \,.
\end{aligned}
\end{equation}
This expression can be written as a single series in $j$; then we can perform the limit $r\rightarrow 1$ on the summand function to obtain the von Neumann entropy
\begin{equation}
\begin{aligned}
S_{\mr{VN}} = &\sum_{j=1}^{\infty} \f{1}{8j \sinh(\tf{1}{2}j\kappa[1+\omega]) \sinh(\tf{1}{2}j\kappa[1-\omega])} \\
&\times \big\{ \coth(\tf{1}{2}j\kappa) \big[2 + j\kappa(1-\omega) \coth(\tf{1}{2}j\kappa[1-\omega]) \\ 
&\kern 7.24em + j\kappa(1+\omega) \coth(\tf{1}{2}j\kappa[1+\omega])\big] \\
&\kern 1.5em + j\kappa \sinh^{-2}(\tf{1}{2}j\kappa) \,\big\} \,.
\end{aligned}
\end{equation}

\subsection{Thermal Wightman functions}
We now focus on the normal-ordered 2-point thermal Wightman function
\begin{multline}
\langle :\!\Phi(x)\,\Phi(x')\!:\rangle_{\mr{LT}} \\ = 2\Re \sum_{n,\ell,m} \langle N_{n,\ell,m}\rangle_{\mr{LT}} \, u_{n,\ell,m}^{\vphantom{*}}(x) \, u^*_{n,\ell,m}(x') \,,
\end{multline}
where
\begin{equation}
\langle N_{n,\ell,m}\rangle_{\mr{LT}} = \f{1}{\exp(\kappa[a\omega_n-\omega m])-1} \,.
\end{equation}
We use the normal-ordering prescription to subtract from the thermal 2-point function the expectation value at zero temperature. This can be easily checked by noticing that $\langle N_{n,\ell,m}\rangle_{\mr{LT}}$ vanishes for $\kappa\rightarrow\infty$.

By using
\begin{equation}
\langle N_{n,\ell,m}\rangle_{\mr{LT}} = \sum_{j=1}^{\infty} \exp(-j\kappa[a\omega_n-\omega m])\,,
\end{equation}
we get
\begin{equation}
\langle :\!\Phi(x)\,\Phi(x')\!:\rangle_{\mr{LT}} = 2\Re \sum_{j=1}^{\infty} \mc{W}(x_j,x') \,,
\end{equation}
where
\begin{equation}
\mc{W}(x,x') = \sum_{n,\ell,m} u_{n,\ell,m}^{\vphantom{*}}(x) \, u^*_{n,\ell,m}(x') \,,
\end{equation}
and $x_j$ is defined by the complex shift
\begin{equation}
t_j = t - ij\kappa a \,, \qquad \varphi_j = \varphi - ij\kappa\omega \,.
\end{equation}
One can sum in $m$ an $\ell$ and express the result in terms of the angle $\delta$ on $\mb{S}^3$, defined with
\begin{multline}
\cos\delta = \cos\chi \cos\chi' + \sin\chi \sin\chi' \\ \times[\cos\theta \cos\theta' + \sin\theta \sin\theta' \cos(\varphi-\varphi')] \,,
\end{multline}
and the adimensionalized time difference $\sigma = a^{-1}(t-t')$ as (see Appendix \ref{APP:summing_polynomials})
\begin{equation}\label{W_general}
\mc W(x,x') = \sum_{n=0}^{\infty} \f{n+1}{4 \pi^2 a^3 \omega_n} e^{-ia\omega_n\sigma} C^1_n(\cos\delta) \,.
\end{equation}

For a conformally coupled field it is easy to perform the sum over $n$ (see again Appendix \ref{APP:summing_polynomials}) which leads to (see also \cite{Camporesi1990}, \cite{Birrell1984} and reference therein)
\begin{equation}
\mc W(x_j,x') = \f{1}{8 \pi^2 a^2(\cos \sigma_j - \cos \delta_j)} \,,
\end{equation}
where $\sigma_j$ and $\delta_j$ are defined by the corresponding shift of $t$ and $\varphi$. Then
\begin{equation}\label{W_conformal}
\langle :\!\Phi(x)\,\Phi(x')\!:\rangle_{\mr{LT}} = \Re \sum_{j=1}^\infty \f{1}{4 \pi^2 a^2 (\cos \sigma_j - \cos \delta_j)} \,.
\end{equation}

Taking the limit for $\sigma,\delta\rightarrow 0$ one obtains
\begin{widetext}
\begin{equation}
\langle :\!\Phi^2(x)\!:\rangle_{\mr{LT}} = \f{1}{4 \pi^2 a^2} \sum_{j=1}^\infty \f{1}{\cosh(j\kappa) - \cos^2\!\chi - \sin^2\!\chi [\cos^2\!\theta + \sin^2\!\theta \cosh(j\kappa\omega)]} \,.
\end{equation}
\end{widetext}
Restoring the dependence of $a$ through $\kappa$ one can see that in the large radius limit the result approaches the asymptotic value
\begin{equation}
\lim_{a\rightarrow\infty} \langle :\!\Phi^2(t,\chi,\theta,\varphi)\!:\rangle_{\mr{LT}} = \tf{1}{12} [T(\chi,\theta)]^2 \,,
\end{equation}
which is precisely the result for homogeneous equilibrium in Minkowski spacetime with temperature $T=T(\chi,\theta)$. Some authors proposed this observable as a ``thermometer'' for a definition of $T$ \cite{Buchholz2002}.

\subsection{Stress-energy tensor}
We want to express the thermal expectation value of \eqref{stress_energy_tensor} in terms of Wightman functions and their derivatives. This is possible adopting the \emph{point-splitting} covariant regularization scheme, through which a composite operator is obtained as the pinching limit of a non-local operator. By using the equation of motion and substituting the values of $M$ and $\xi$ for a conformally coupled field we get
\begin{widetext}
\begin{equation}
\langle :\!T_{\mu\nu}(x)\!:\rangle_{\mr{LT}} = \tf{1}{6}\lim_{x'\rightarrow x}\big\{ 4\nabla_\mu{\nabla'}_{\!\nu} - \nabla_{\mu}\nabla_{\nu} - {\nabla'}_{\!\mu}{\nabla'}_{\!\nu} -g_{\mu\nu} \big(g^{\rho\sigma} \nabla_\rho{\nabla'}_{\!\sigma}-a^{-2}\big)-R_{\mu\nu}\big\} \, \langle :\!\Phi(x)\,\Phi(x')\!:\rangle_{\mr{LT}} \,.
\end{equation}
\end{widetext}

As pointed out earlier, we introduce normal ordering to compute the statistical average of the stress-energy tensor as the shift from the zero-temperature value. The result for $T \rightarrow 0$ corresponds to the expectation value on the static vacuum, i.e.\ the vacuum associated to a static observer and selected by our choice of coordinates. The problem of deriving a regularized expression for the vacuum term has been originally discussed in \cite{Ford1976}.

For $\omega=0$ we obtain the thermal expectation value of the stress-energy tensor in the special case of homogeneous equilibrium \cite{Dowker1977, Dowker1978}, that can be written in terms of a $q$-polygamma function as
\begin{equation}
\begin{aligned}
\langle :\!T_{tt}\!:\rangle &= \f{\psi^{(3)}_{e^\kappa}(1)}{2 \pi^2 (a\kappa)^4} \,, \\
\langle :\!T_{\varphi\varphi}\!:\rangle &= \sin^2\!\theta \, \langle :\!T_{\theta\theta}\!:\rangle \\
  &= \sin^2\!\chi \sin^2\!\theta \, \langle :\!T_{\chi\chi}\!:\rangle \\
  &= \tf{1}{3} a^2 \sin^2\!\chi \sin^2\!\theta \, \langle :\!T_{tt}\!:\rangle \,.
\end{aligned}
\end{equation}

In the general case, the components of the stress-energy tensor exhibit a rather complex expression. Let us consider the energy density
\begin{equation}
\epsilon = u^{\mu} u^{\nu} \langle :\!T_{\mu\nu}(x)\!:\rangle_{\mr{LT}} \,.
\end{equation}
As mentioned above, symmetries ensure that the result will depend on $x$ only through $\alpha$. In particular, we can write the result in terms of $S=\sin^2\!\alpha$ as
\begin{multline}
\epsilon = \sum_{j=1}^{\infty} \f{\mc{N}}{12 \pi^2 a^4 (1-S\omega^2)} \\
  \times \big[1 - S - \cosh(j\kappa) + S\cosh(j\kappa\omega)\big]^{-3} \,,
\end{multline}
where
\begin{equation}
\begin{aligned}
\mc{N} = \mc{A} &+ \mc{B} \cosh(j\kappa) + \mc{C} \cosh(j\kappa\omega) \\
  &+ \mc{D} \cosh(j\kappa) \cosh(j\kappa\omega) \\
  &+ \mc{E} \sinh^2(j\kappa) + \mc{F} \sinh^2(j\kappa\omega) \\
  &+ \mc{G} \sinh(j\kappa) \sinh(j\kappa\omega)
\end{aligned}
\end{equation}
and
\begin{align}
\mc{A} &= 2(1+S^2\omega^2) \,, \nonumber \\
\mc{B} &= -2+S+S^2\omega^2 \,, \nonumber \\
\mc{C} &= 1+S\omega^2-2S^2\omega^2 \,, \nonumber \\
\mc{D} &= -(1+S+S\omega^2+S^2\omega^2) \,, \\
\mc{E} &= -3-S\omega^2 \,, \nonumber \\
\mc{F} &= -S(1+3S\omega^2) \,, \nonumber \\
\mc{G} &= 12S\omega \,. \nonumber
\end{align}

It is possible to characterize the asymptotic behavior of the energy density for large radius performing an expansion in $a^{-1}$. This gives
\begin{equation}\label{epsilon_expansion}
\epsilon \sim \f{\pi^2T^4}{30} + \f{(1-S) \omega^2 T^2}{36 (1-S\omega^2)^2} \, a^{-2} + o(a^{-2}) \,,
\end{equation}
where again $T$ is the local proper temperature, function of $S$ and $\omega$.
The leading order is the homogeneous equilibrium result in Minkowski spacetime; this is expected, since when $a$ diverges, both gradients of $\beta$ and curvature go to zero, and finite-volume effects vanish.

To better understand the origin of the subleading term we define the \emph{vorticity} vector field
\begin{equation}
\Omega^\mu = \tf{1}{2} \epsilon^{\mu\nu\rho\sigma} u_\nu \nabla_\rho u_\sigma \,;
\end{equation}
using \eqref{u_field} one finds
\begin{equation}
\Omega^2 = -\f{(1-S) \omega^2}{a^2 (1-S\omega^2)^2} \,.
\end{equation}
We can then rewrite \eqref{epsilon_expansion} as
\begin{equation}
\epsilon \sim \tf{1}{30} \pi^2 T^4 - \tf{1}{36} \Omega^2 T^2 + o(a^{-2}) \,,
\end{equation}
which is precisely the result obtained performing an hydrodynamic expansion of $\epsilon$ in flat space in terms of the gradients of the $\beta$ field \cite{Becattini2015}. This seems to indicate that, in the limit of large radius, finite-volume effects decay faster than effects associated with thermal gradients. 

Now, since $u$, $A$, and $\Omega$ are mutually orthogonal, we can introduce a fourth vector field 
\begin{equation}
Q^\mu = \epsilon^{\rho\sigma\tau\mu} u_\rho A_\sigma \Omega_\tau
\end{equation}
to form an orthogonal tetrad. By contracting $\langle :\!T_{\mu\nu}\!:\rangle_{\mr{LT}}$ with elements of the tetrad we see that the only non-vanishing ``mixed'' contraction is given by $u^\mu Q^\nu \langle :\!T_{\mu\nu}\!:\rangle_{\mr{LT}}$ (together with the symmetric term). Then, the expectation value of the stress energy tensor can be written as
\begin{multline}\label{T_decomposition}
\langle :\!T_{\mu\nu}\!:\rangle_{\mr{LT}} = \epsilon \, u_\mu u_\nu + q \, (u_\mu \h{Q}_\nu + \h{Q}_\mu u_\nu) \\
  + p_A \, \h{A}_\mu \h{A}_\nu + p_\Omega \, \h{\Omega}_\mu \h{\Omega}_\nu + p_Q \, \h{Q}_\mu \h{Q}_\nu \,,
\end{multline}
where we introduced the normalization of four-vectors as $\h{V}^\mu = |V^2|^{-\f{1}{2}} V^\mu$. The various coefficients that appear in the previous expression can be found by performing appropriate contractions. The different values of $p_A$, $p_\Omega$, and $p_Q$ are associated with pressure anisotropy, and turn out to be 
\begin{align}
p_A &= \sum_{j=1}^{\infty} \f{1 + \cosh(j\kappa) + \cosh(j\kappa\omega)}{12 \pi^2 a^4 [1 - S - \cosh(j\kappa) + S\cosh(j\kappa\omega)]^{2}} \,, \nonumber \\
p_\Omega &= p_A + \sum_{j=1}^{\infty} \f{2 (1-S) \sinh^2(\tf{1}{2}j\kappa) \sinh^2(\tf{1}{2}j\kappa\omega)}{3 \pi^2 a^4 [1 - S - \cosh(j\kappa) + S\cosh(j\kappa\omega)]^{3}} \,, \nonumber \\
p_Q &= \epsilon - p_A - p_\Omega \,,
\end{align}
where for the last identity we used the fact that, for a conformally coupled field, the trace of the stress-energy tensor vanishes. Finally, the coefficient of the transverse term reads
\begin{multline}
q = \sum_{j=1}^{\infty} \f{\s{S} \mc{N}'}{6 \pi^2 a^4 (1-S\omega^2)} \\
  \times \big[1 - S - \cosh(j\kappa) + S\cosh(j\kappa\omega)\big]^{-3} \,,
\end{multline}
where $\mc{N}'$ is defined analogously to $\mc{N}$, but now we have
\begin{align}
\mc{A}' &= -(1+S)\omega \,, \nonumber \\
\mc{B}' &= (1-S)\omega \,, \nonumber \\
\mc{C}' &= -(1-S)\omega \,, \nonumber \\
\mc{D}' &= (1+S)\omega \,, \\
\mc{E}' &= \omega \,, \nonumber \\
\mc{F}' &= S\omega \,, \nonumber \\
\mc{G}' &= -3(1+S\omega^2) \,. \nonumber
\end{align}

In the limit of large radius one finds
\begin{align}
p_A \sim p_Q &\sim \tf{1}{90} \pi^2 T^4 - \tf{1}{36} \Omega^2 T^2 + o(a^{-2}) \,, \nonumber \\
p_\Omega &\sim \tf{1}{90} \pi^2 T^4 + \tf{1}{36} \Omega^2 T^2 + o(a^{-2}) \,,
\end{align}
which, again, are in agreement with \cite{Becattini2015}. That is not the case, however, for
\begin{equation}
q \sim -\f{\s{S} \omega (1-\omega^2)}{18 a^2 (1-S\omega^2)^2} T^2 + o(a^{-2}) \,,
\end{equation}
where, apparently, other effects intervene that cannot be captured by an hydrodynamic expansion in Minkowski spacetime.

\section{Discussion}\label{SEC:discussions}
In this paper we have studied a class of equilibrium states on stationary spacetimes through a model-independent approach based on a geometric definition of local thermodynamic variables. We have applied this framework to the case of a scalar field in the Einstein's static universe and we have derived the associated partition function and thermal expectation value of the stress energy tensor.

As it also appears from the results, because thermodynamic variables have been introduced without a direct connection with the expectation value of some local operator, in general there is no obvious way on how to express the latter in terms of the former without resorting to some particular limit. This is a signal of the fact that, in general, expectation values of local operators depend on thermodynamic fields in some non-local way.

A more familiar condition, however, is realized for large values of the radius $a$, where a hydrodynamic limit for $\langle :\!T_{\mu\nu}\!:\rangle_{\mr{LT}}$ is recovered, as results exhibit a local dependence on $\beta$ and, at subleading orders, on its gradients. An even clearer interpretation can be obtained noticing that performing the limit for large $a$ is the same of performing the limit for large $\tau$ (as they both appear through $\kappa$) if not for the presence of an additional $a^{-4}$ factor which ensures the convergence in the thermodynamic limit, while a $\tau^4$ scaling characterizes the high-temperature limit. Nonetheless the picture is clear: at high temperatures thermal correlation lengths become small and the components of $\langle :\!T_{\mu\nu}\!:\rangle_{\mr{LT}}$ depend only on the value of thermodynamic fields at nearby points.

However, considering the opposite limit can be equally interesting; let us see why. One of the key features of this model is the fact that it is well defined even on a global scale. In particular, field quantization can be carried out on the whole spacetime, and the Killing vector fields that define equilibrium states are global and, more importantly, globally time-like. These two features constitute a strong constraint and restrict the number of non-trivial examples of this kind that one can find. As a consequence, this model can be particularly useful to probe the low temperature regime where thermal correlations capture long-distance---or even global---properties of the thermal state and of the background geometry. This could also allow, in principle, to discard spurious effects that may arise in models whose thermal interpretation cannot be extended to a global scale.

\section*{Acknowledgements}
It is a pleasure to thank \mbox{J. Hayling}, \mbox{M. Poggi}, and \mbox{R. Russo} for reading the manuscript and providing suggestions, as well as \mbox{E. Grossi}, \mbox{N. Pinamonti}, and \mbox{D. Seminara} for interesting discussions.

\appendix

\section{Orthonormality of KG solutions}\label{APP:orthonormality}
Gegenbauer polynomials (also called ultraspherical polynomials) constitute a generalization of Legendre polynomials. They satisfy the orthonormality relation \cite{Abramowitz1964}
\begin{equation}
\int_{-1}^1 \! \d x \, (1-x^2)^{\alpha-\f{1}{2}} \, C^\alpha_r(x) \, C^\alpha_s(x) = \f{\pi 2^{1-2\alpha} \Gamma(r+2\alpha)}{r!(r+\alpha)[\Gamma(\alpha)]^2} \, \delta_{r,s} \,,
\end{equation}
which, in turn, ensures that
\begin{equation}
\int_0^\pi \d\chi \, \sin^2\!\chi \, \Pi^\ell_n(\chi) \, \Pi^{\ell}_{n'}(\chi) = \delta_{n,n'} \,.
\end{equation}
This, together with the well-known properties of spherical harmonics, is sufficient to determine the orthonormality of the solutions $u_{n,\ell,m}$. In fact, it is easy to check that these satisfy the required relation
\begin{multline}
-i\int_{\sigma} \d\sigma \, n^\mu \l(u_{n,\ell,m}^{\vphantom{*}} \partial_\mu u_{n',\ell',m'}^* - \partial_\mu u_{n,\ell,m}^{\vphantom{*}} u_{n',\ell',m'}^*\r) \\
  = \delta_{n,n'}\delta_{\ell,\ell'}\delta_{m,m'} \,.
\end{multline}

\section{Summing ultraspherical polynomials}\label{APP:summing_polynomials}
Legendre polynomials can be recovered from Gegenbauer ones setting the superscript of the latter to $\f{1}{2}$; thus, the spherical harmonics addition theorem reads
\begin{equation}
\sum_{m=-\ell}^{\ell} Y^m_\ell(\theta,\varphi) \, Y^{m*}_\ell(\theta',\varphi') = \f{2\ell+1}{4\pi} \, C^{\f{1}{2}}_\ell(\cos \gamma) \,,
\end{equation}
where $\cos \gamma = \cos \theta \cos \theta' + \sin \theta \sin \theta' \cos (\varphi-\varphi') \,.$
We will also make use of
\begin{multline}
\sum_{\ell=0}^n \f{2^{2\ell}(2\ell+1)(n-\ell)![\ell!]^2}{(n+\ell+1)!} \, (\sin \chi)^\ell \,C^{1+\ell}_{n-\ell}(\cos \chi) \\ \times(\sin \chi')^\ell \, C^{1+\ell}_{n-\ell}(\cos \chi') \, C^{\f{1}{2}}_\ell(\cos \gamma) = C^1_n(\cos\delta) \,, \\
\end{multline}
where $\cos\delta = \cos\chi \cos\chi' + \sin\chi \sin\chi' \cos\gamma$ (see \cite{Bateman1953}, eq.~10.9(34); there, the misprint $2^n$ should read $2^{2n}$).

Combining the last two identities, it is straightforward to derive \eqref{W_general}.
In the conformal case we can apply \cite{Abramowitz1964}
\begin{equation}
\sum_{n=0}^\infty t^n \, C_n^\lambda(x) = \f{1}{(1-2xt+t^2)^\lambda} \,, \qquad |t|<1\,, \quad |x|\leq 1 \,.
\end{equation}
In fact, from
\begin{equation}
\mc W(x,x') = \f{e^{-i\sigma}}{4\pi^2a^2} \sum_{n=0}^{\infty} [e^{-i\sigma}]^n \, C^1_n(\cos\delta) \,,
\end{equation}
\eqref{W_conformal} follows immediately.

\section{Residual symmetries}\label{APP:residual_symmetries}
The orbits of the integral curves of $r$ can be characterized as contour lines of some function $f(\chi,\theta)$. Such a function obeys the equation
\begin{equation}
-\cos\theta \, \partial_\chi f + \cot\chi \sin\theta \, \partial_\theta f = 0 \,,
\end{equation}
that can be solved by separation of variables, leading to
\begin{equation}
f \propto \sin\chi \sin\theta
\end{equation}
up to an additive constant. This result is certainly not unexpected, since $f$ is precisely the function that has to remain constant if we want $T(\chi,\theta)$ to preserve its value. When $\theta$ is rotated to $\f{\pi}{2}$, we obtain equation \eqref{alpha_eq} for $\alpha$.

\bibliography{bibliography}

\end{document}